**Comparative Analysis of Brookhaven National Laboratory Nuclear Decay Data and Super-Kamiokande Neutrino Data: Indication of a Solar Connection**

P.A. Sturrock[a,*], E. Fischbach[b]

[a] The Kavli Institute for Particle Astrophysics and Cosmology and the Center for Space Science and Astrophysics, Stanford University, Stanford, CA 94305-4060, USA.
[b] Department of Physics and Astronomy, Purdue University, West Lafayette, IN 47907, USA.

*Corresponding author. Tel +1 6507231438; fax +1 6507234840.
Email address: sturrock@stanford.edu

ABSTRACT An experiment carried out at the Brookhaven National Laboratory from February 1982 to December 1989 acquired 364 measurements of the beta-decay rates of a sample of $^{36}$Cl and of a sample of $^{32}$Si. The experimenters reported finding *small periodic annual deviations of the data points from an exponential decay… of uncertain origin*. We here analyze this dataset by power spectrum analysis and by forming spectrograms and phasegrams. We confirm the occurrence of annual oscillations but we also find evidence of oscillations in a band of frequencies appropriate for the internal rotation of the Sun. Both datasets show clear evidence of a transient oscillation of frequency 12.7 year$^{-1}$, which falls in the range of rotational frequencies for the solar radiative zone. We repeat these analyses for 358 neutrino measurements acquired by Super-Kamiokande over the interval May 1986 to August 2001. Spectrogram analysis yields a strong and steady oscillation at about 9.5 year$^{-1}$ and an intermittent oscillation with frequency in the range 12.5 – 12.7 year$^{-1}$. We attribute the former to rotation of the solar core and the latter to rotation in the radiative zone. Since the flux of neutrinos ($^8$B neutrinos) responsible for the Super-Kamiokande measurements is known, we are able to estimate the cross sections for the beta-decay oscillations at 12.7 year$^{-1}$. These estimates are found to be $10^{-21.6}$ cm$^{-2}$ for $^{36}$Cl and $10^{-18.4}$ cm$^{-2}$ for $^{32}$Si. We suggest that the beta-decay process is influenced by neutrinos due to some currently unknown mechanism and that the solar neutrino flux is modulated by magnetic field in the deep solar interior by Resonant Spin Flavor Precession.

**Section 1. Introduction**
There has long been interest in data acquired at the Brookhaven National Laboratory (BNL) from an experiment designed to determine the half-life of $^{32}$Si (Alburger *et al.*, 1986). For almost 8 years, the experimenters tracked the (beta) decay rate of a specimen of $^{32}$Si and, for comparison, the decay rate of a specimen of the long-lived nuclide $^{36}$Cl (half life approximately 300,000 years). Measurements beginning in February 1982 were made at approximately 4–week intervals, each consisting of a total of 40 hours of counting for each specimen. For each nuclide, the daily decay rate was formed from an average of 20 separate measurements in the course of the day, the $^{36}$Cl and $^{32}$Si measurements being interwoven. The interest arises from the fact that the authors reported that *small periodic annual deviations of the data points from an exponential decay were observed, but were of uncertain origin*. These oscillations were found to have an amplitude (depth of modulation) of $6 \times 10^{-4}$, with a maximum at February 9. Other investigators have independently noted evidence of oscillations or other variations in nuclear decay rates (e.g. Baurov, 2010; Falkenberg, 2001; Parkhomov, 2011).

The BNL experiment attracted the attention of Jenkins *et al.* (2009), who speculated that the annual oscillation might be related to the annual variation of the Earth-Sun distance (as had been proposed by Falkenberg (2001)). Examining the ratio of the $^{32}$Si and $^{36}$Cl decay rates, Jenkins *et al.* confirmed that the maximum occurs in early February, not far from the date (January 3) of closest approach of Earth to the Sun. They noted that long-term measurements of the decay rate of $^{322}$Ra, carried out at the Physikalisch Technische Bundesanstalt (Siegert *et al.*,1998), appeared to show a similar annual oscillation. Jenkins *et al.* suggested that solar neutrinos may be responsible for variations in beta-decay rates. This suggestion was of course viewed with great suspicion, since the cross section for collisions between neutrinos and known particles is believed to be of order $10^{-43}$ cm$^2$ (Bahcall, 1989).

Power spectrum analysis of the BNL data also gave evidence of periodicities with frequencies in the range 11 – 13 year$^{-1}$, corresponding to periods approximately in the range 25 – 35 days (Javorsek *et al.,* 2010; Sturrock, Buncher, *et al.*, 2010). Oscillations of decay data with frequencies in this range had previously been identified by Parkhomov (2011) in studies of the decay rate of several nuclides. These results, if valid, would support the proposal of a solar influence, since this frequency range is consistent with estimates of the solar rotation frequency (that is known to be a function of radius and latitude; Schou *et al.,* 2002).

However, the results of these investigations have not been entirely consistent, which has led to the suspicion that different nuclides may have different decay properties, and to the possibility that these properties may be variable – in particular, that oscillations may be intermittent. For instance some recent experiments, that were designed to give highly accurate measurements of decay rates, have yielded no evidence of variability (Bellotti *et al*., 2012; Kossert *et al*., 2014, 2015). The lack of evidence of variability may indicate that the decay rate of the nuclide under investigation is in fact constant. However, it may be attributable to the possibility that the decay is oscillatory but the oscillations are intermittent and/or that evidence of variability may depend on details of the measurement process such as sensitivity to different energy ranges of the daughter products.

The purpose of this article is to examine the possibility that the decay rates of some nuclides may be oscillatory with a frequency compatible with that of solar rotation, but the oscillations may be intermittent or vary in some other way. We investigate this possibility by means of spectrogram analysis of the BNL dataset.

**Section 2. BNL power spectrum analysis**
The BNL dataset comprises 364 daily decay-rate measurements of each of $^{36}$Cl and $^{32}$Si acquired over the interval 1982.106 to 1989.929 (duration 7.823 years; Alburger *et al.,* 1986). For each nuclide, the daily decay rate was formed from an average of 20 separate measurements in the course of the day, the $^{36}$Cl and $^{32}$Si measurements being interwoven. Plots of the daily count rates (counts per 600 min.) are shown in Figures 1 and 2.

For convenience of power spectrum analysis, we adopt a date format (one that does not involve leap years) that has proved useful in the analysis of solar neutrino data (Sturrock, 2006). We first count dates in "neutrino days," for which January 1, 1970, is designated "neutrino day 1" ($t(ND) = 1$). We then convert dates to "neutrino years", denoted by $t(NY),$ as follows:

$$t(NY) = 1970 + t(ND)/365.2564. \tag{1}$$

Dates in neutrino years differ from true dates by no more than one day.

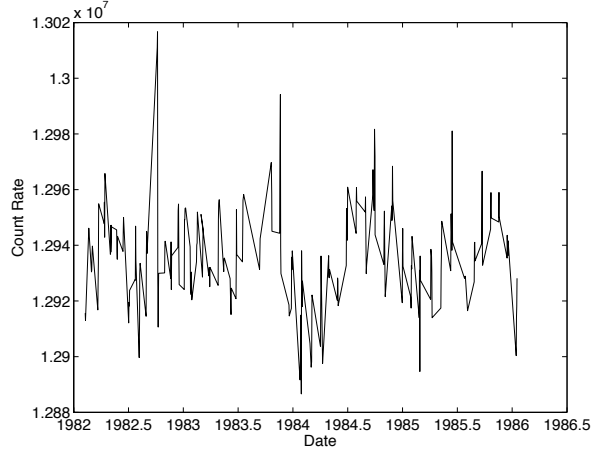

Figure 1. Plot of the $^{36}$Cl count rate (number of counts per 600 minutes).

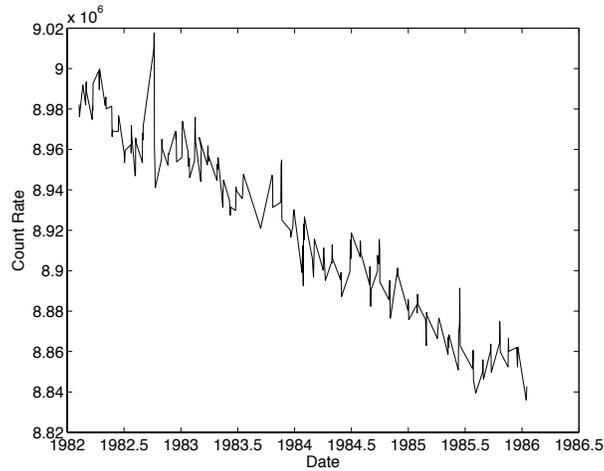

Figure 2. Plot of the $^{32}$Si count rate (number of counts per 600 minutes).

We next detrend (to remove the exponential decay) and normalize the data for each nuclide as follows. From the times $t_n$ and count-rate measurements $y_n$, we form

$$V = \sum_n \left(\log(x_n) + \kappa t_n - C\right)^2 \tag{2}$$

and determine the values of $\kappa$ and $K$ that minimize $V$. The normalized (and detrended) values are then given by

$$z_n = x_n/y_n \text{ where } y_n = \exp(C - \kappa t_n). \tag{3}$$

Plots formed from the normalized decay-rates for $^{36}$Cl and $^{32}$Si are shown in Figures 3 and 4, respectively. It is interesting that, for $^{36}$Cl, the standard deviation of measurements for 1986 to 1990 is notably less than that of measurements from 1982 to 1986. For $^{32}$Si, measurements for 1985 to 1987 are a little lower than measurements for the rest of the dataset; measurements for $^{36}$Cl do not show a similar feature. These comparisons suggest that $^{36}$Cl and $^{32}$Si were not subject to exactly the same influences, despite the fact that their environmental conditions were identical.

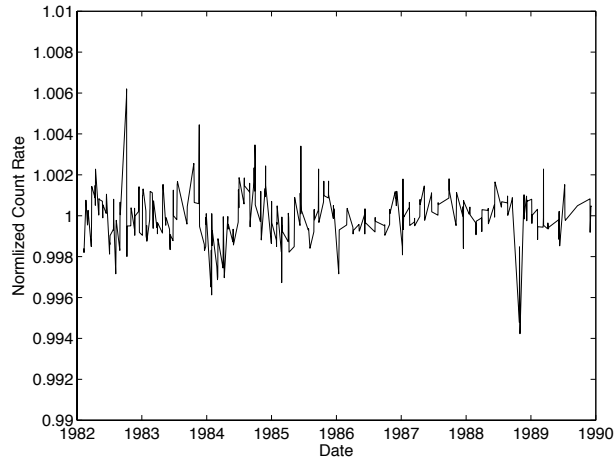

Figure 3. A plot of detrended and normalized $^{36}$Cl data.

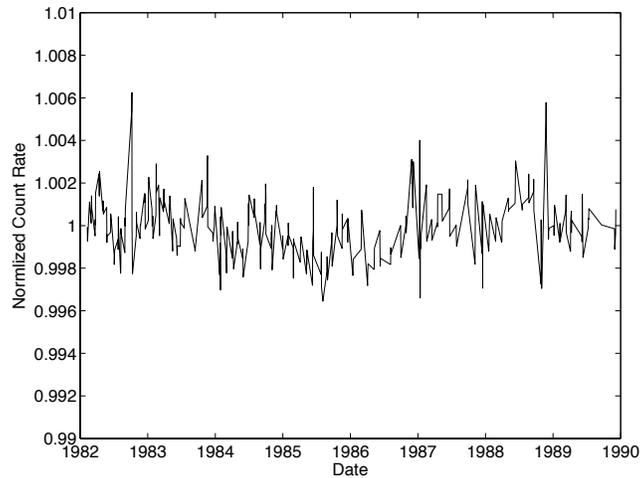

Figure 4. A plot of detrended and normalized $^{32}$Si data .

We have carried out power-spectrum analyses for the frequency range $\nu = 0 - 16$ year$^{-1}$ of both datasets, using a likelihood procedure (Sturrock, 2003). The results are shown in Figure 5 and 6, and the top 20 peaks are shown in Tables 1 and 2.

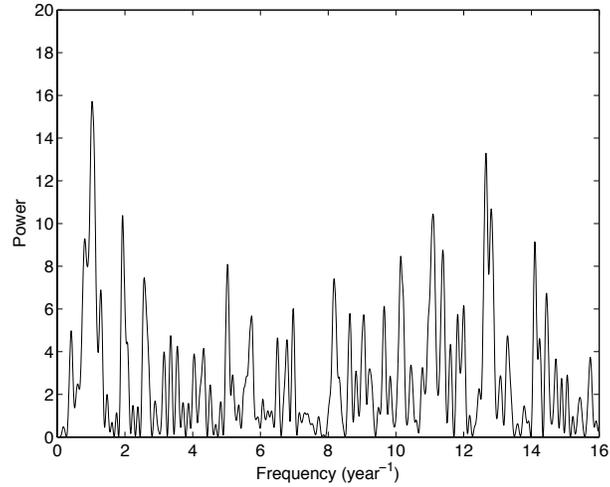

Figure 5. Power spectrum formed from normalized $^{36}$Cl measurements.

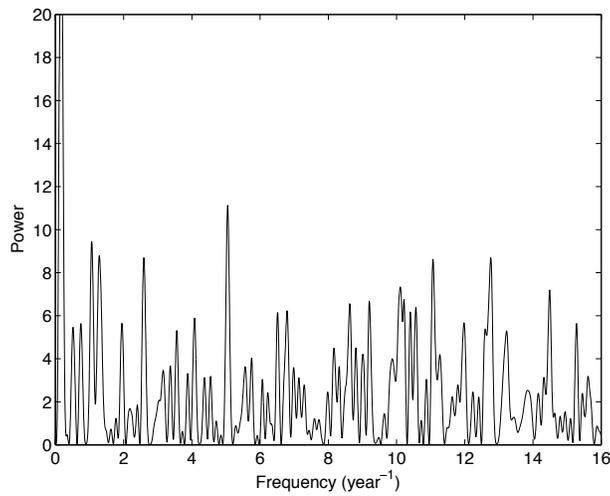

Figure 6. Power spectrum formed from normalized $^{32}$Si measurements.

We see that the power spectrum for $^{36}$Cl shows a very strong annual modulation (power 15.71 at frequency 1.3 year$^{-1}$). The power at exactly 1.00 year$^{-1}$. is 14.61. For an expected exponential power distribution, the probability of finding this value or more by chance is $e^{-S}$, i.e. 4.5 10$^{-7}$ (Scargle, 1982).

Table 1. Top 20 peaks in the frequency band 0 – 16 year$^{-1}$ in the power spectrum formed from the decay-rate measurements for $^{36}$Cl.

| Frequency (year$^{-1}$) | Power | Order |
|---|---|---|
| 0.82 | 9.28 | 6 |
| 1.03 | 15.71 | 1 |
| 1.29 | 6.89 | 13 |
| 1.93 | 10.37 | 5 |
| 2.57 | 7.46 | 11 |
| 5.03 | 8.09 | 10 |
| 6.97 | 6.01 | 17 |
| 8.17 | 7.41 | 12 |
| 8.64 | 5.79 | 18 |
| 9.05 | 5.73 | 20 |
| 9.65 | 6.13 | 16 |
| 10.14 | 8.47 | 9 |
| 11.10 | 10.44 | 4 |
| 11.39 | 8.75 | 8 |
| 11.82 | 5.74 | 19 |
| 12.00 | 6.17 | 15 |
| 12.66 | 13.29 | 2 |
| 12.81 | 10.68 | 3 |
| 14.10 | 9.14 | 7 |
| 14.45 | 6.74 | 14 |

Table 2. Top 20 peaks in the frequency band 0 – 16 year$^{-1}$ in the power spectrum formed from the decay-rate measurements for $^{32}$Si.

| Frequency (year$^{-1}$) | Power | Order |
|---|---|---|
| 0.17 | 22.84 | 1 |
| 1.06 | 9.45 | 3 |
| 1.29 | 8.79 | 4 |
| 1.95 | 5.64 | 19 |
| 2.59 | 8.70 | 5 |
| 4.08 | 5.89 | 17 |
| 5.05 | 11.13 | 2 |
| 6.51 | 6.14 | 16 |
| 6.79 | 6.23 | 14 |
| 8.63 | 6.56 | 12 |
| 9.20 | 6.68 | 11 |
| 10.11 | 7.34 | 8 |
| 10.22 | 6.76 | 10 |
| 10.41 | 6.17 | 15 |
| 10.57 | 6.38 | 13 |
| 11.07 | 8.63 | 7 |
| 11.98 | 5.67 | 18 |
| 12.76 | 8.69 | 6 |
| 14.49 | 7.20 | 9 |
| 15.28 | 5.64 | 20 |

We have carried out a shuffle test (Bahcall *et al.,* 1991) of the $^{36}$Cl data, randomly re-assigning values $x_n$ to values $t_n$. The result of 10,000 shuffles is shown in Figure 7. According to this test, the probability of finding the power 14.61 or more by chance is 2.7 10$^{-7}$. Both tests show that the probability of finding by chance the annual modulation shown in Figure 5 is less than 5 10$^{-7}$.

It is interesting that power-spectrum analysis of the $^{32}$Si data, shown in Figure 6, does not show such a strong annual modulation. We see from Table 2 that there is a peak with power 9.45 at frequency 1.06 year$^{-1}$. The power at exactly 1.00 year$^{-1}$ is 3.51, for which e$^{-S}$ has the value 0.02. It is notable that the $^{36}$Cl and $^{32}$Si power spectra are quite different, even though the environmental conditions were identical.

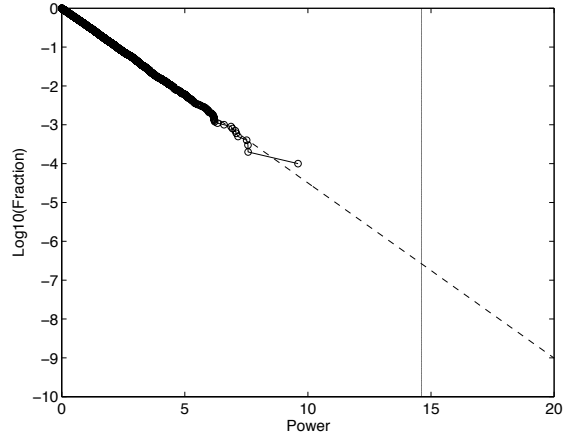

Figure 7. Results of a shuffle test of the annual modulation in $^{36}$Cl data. There is a probability of only 3 10$^{-7}$ of obtaining by chance the actual power 14.71 at the frequency 1.00 year$^{-1}$.

It is interesting to note from Tables 1 and 2 that there is some evidence for harmonics of the annual modulation, notably oscillations with frequencies close to 2 year$^{-1}$ and 5 year$^{-1}$. Harmonics, if real, would not be compatible with the early suggestion that an annual modulation is due to the varying Earth-Sun distance (Falkenberg, 2001, Jenkins *et al.,* 2009).

Table 3 shows the power, amplitude (depth of modulation) and phase (phase of maximum always intended) of the annual oscillation (at exactly 1.00 year$^{-1}$) for each nuclide, as derived from our count-rate analysis. We obtain an estimate of 171.5 years for $^{32}$Si that is consistent with that derived by Alburger *et al.* (1986). The half-life of $^{36}$Cl is too long to determine in just a few years.

Table 3. Inferred half-life and power, amplitude and phase of maximum of the annual modulation ($\nu = 1$ year$^{-1}$) of the normalized $^{36}$Cl and $^{32}$Si data.
(The $^{36}$Cl half-life is too long to determine accurately.)

| Nuclide | Half-life (y) | Power | Amplitude | Phase of Maximum |
|---|---|---|---|---|
| $^{36}$Cl | | 14.46 | 5.29e-04 | 0.70 |
| $^{32}$Si | 171.5 | 3.52 | 2.72e-04 | 0.90 |

**Section 3. BNL spectrogram and phasegram analyses of the annual modulations**

We see from Figure 3 that the $^{36}$Cl dataset is distinctly non-uniform, and there is some evidence from Figure 4 that the same is true of the $^{32}$Si dataset. It is therefore helpful to examine these datasets by means of spectrograms and phasegrams. Figures 8 and 9 show spectrograms formed from $^{36}$Cl data and $^{32}$Si data, respectively, for the frequency band 0 – 8 year$^{-1}$. Each spectrogram is formed by selecting sequentially 200 consecutive measurements and forming the power of that selection, using the same likelihood procedure as in Section 2 (Sturrock, 2003). The power is indicated by the color.

We see from Figure 8 that $^{36}$Cl data show a strong annual modulation for the time interval 1984 to 1986. It appears that the reduction in standard deviation in the $^{36}$Cl measurements over the second half of the duration corresponds to (and may be caused by) the absence (or reduction) of an annual oscillation over that time frame. We see from Figure 9 that the $^{32}$Si data show intermittent oscillations (but no stable oscillation) near the annual frequency.

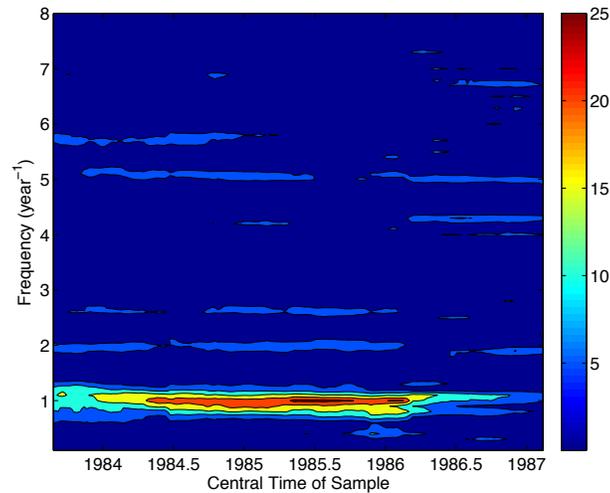

Figure 8. Spectrogram formed from $^{36}$Cl data for the frequency band 0 – 8 year$^{-1}$.

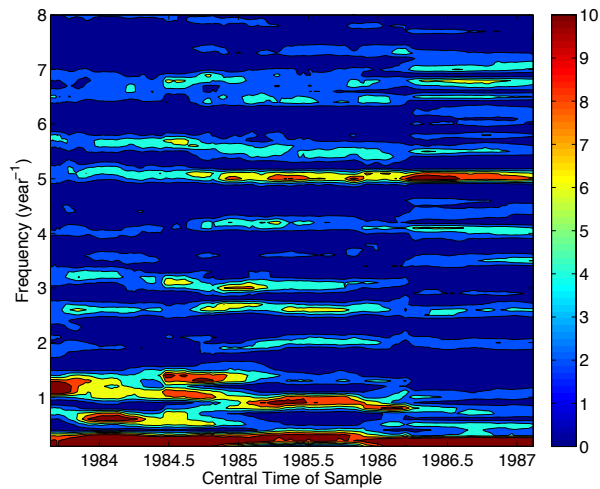

Figure 9. Spectrogram formed from $^{32}$Si data for the frequency band 0 – 8 year$^{-1}$.

We form phasegrams by a procedure similar to that used to form spectrograms. Each phasegram is formed by selecting sequentially 200 consecutive measurements and forming the power of that selection as a function of phase of year (over the range 0 to 1) at the set frequency 1 year$^{-1}$. Figures 10 and 11 show phasegrams formed from the normalized $^{36}$Cl and $^{32}$Si data, respectively.

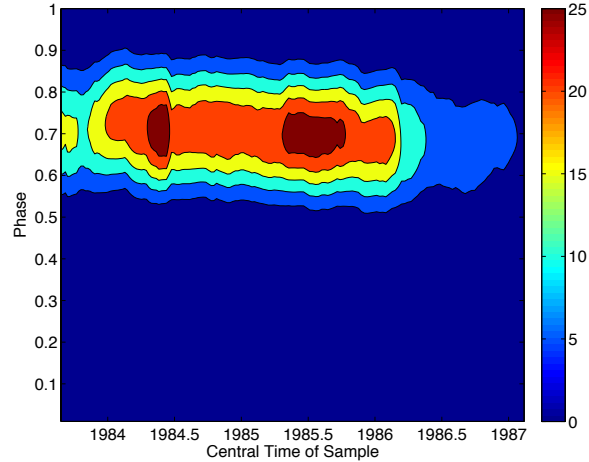

Figure 10. Phasegram formed from $^{36}$Cl data for the annual modulation at frequency 1 year$^{-1}$

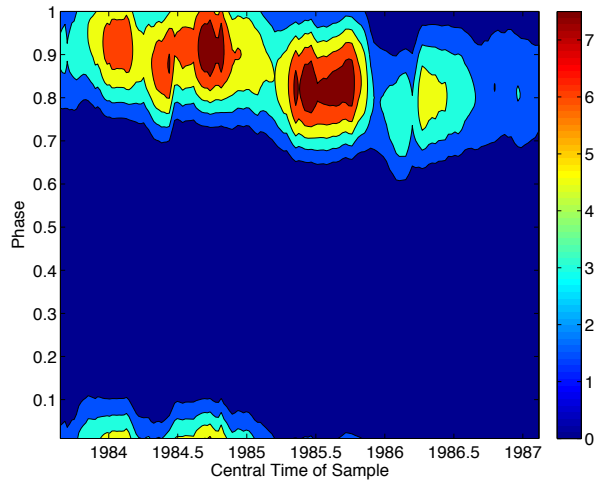

Figure 11. Phasegram formed from $^{32}$Si data for the annual modulation at frequency 1 year$^{-1}$

We see from Figures 8 and 10 that, for $^{36}$Cl, the annual modulation is most evident for the time interval 1984 to 1986, with a phase of 0.71. We see from Figures 9 and 11 that, for $^{32}$Si, the annual modulation is most evident for the same time interval but with a phase that drifts from 0.95 to 0.85.

It is interesting that the two phasegrams are not the same. It is also notable that, for each nuclide, the indicated phase of the decay rate is not close to 0 (or 1), the value one would expect if the oscillations were due simply to the varying Earth-Sun distance.

It appears that, whatever cause (which may be multiple) is responsible for annual variations of beta-decay rates, it is not steady and it does not have exactly the same influence at all times and for all nuclides. This already suggests that the variability of beta-decay rates is not to be understood purely in terms of environmental influences, or in terms of the varying Earth-Sun distance.

**Section 4. BNL spectrogram analysis of possible solar rotational modulations**

As we point out in a recent article (Sturrock, Fischbach, *et al.,* 2015), an annual oscillation of a beta-decay rate may be due to a solar influence, but it may also be due (entirely or in part) to some other influence, such as an annual variation of the environmental conditions or to a cosmic influence. Hence, to test for a solar influence, we should look for other evidence. We here search for an oscillation or oscillations that may be associated with solar rotation. If the influence is due to neutrinos, modulation of the neutrino flux may in principle occur anywhere inside the Sun. Theoretically, the neutrino flux could be influenced by either the MSW (Mikheyev, Smirnov, Wolfenstein) effect (Mikheyev et *al.,* 1986; Wolfenstien, 1978) that is determined by the density structure, or by the RSFP (Resonant Spin Flavor Precession) effect (Akhmedov, 1988) that depends on both the density and the magnetic field. Either process provides a mechanism for converting an electron neutrino (the type produced by nuclear reactions in the core) into either a muon neutrino or a tau neutrino, neither of which would be detected by a neutrino experiment such as Super-Kamiokande (Bahcall, 1989). The fact that neutrino experiments detect fewer neutrinos than expected is attributed to these processes (Ianni, 2014). It seems reasonable to assume that if beta decays are influenced by neutrinos, that influence is due primarily – perhaps exclusively - to electron neutrinos.

The Sun is sufficiently stable that the MSW effect would not lead to any detectable time variation. On the other hand, the solar magnetic field, as it is observed at the photosphere, is highly asymmetric and highly variable. The same is likely to be true throughout the convection zone (normalized radius 0.7 to 1) and may also be true in the radiative zone (normalized radius 0.3 or 0.4 to 0.7). Hence if the RSFP process is operative in a region where the magnetic field is sufficiently strong and sufficiently inhomogeneous, we may expect that the solar neutrino flux may exhibit modulation in a band of frequencies appropriate to the Sun's internal rotation. As determined by helioseismology, the equatorial sidereal rotation rate in the radiative zone is in the range 13.5 – 15.0 year$^{-1}$, which converts to a synodic rate (as seen from Earth) of 12.5 – 14.0 year$^{-1}$ (Schou *et al*., 2002). However, the rotation rate in the deep interior is quite uncertain and there are indications, from analyses of Super-Kamiokande measurements (Sturrock & Scargle*,* 2006), that some part of the solar interior may rotate as slowly as 10.4 year$^{-1}$ (sidereal) or 9.4 year$^{-1}$ (synodic). Hence a reasonable search band for the rotational modulation of beta-decay rates (as seen from Earth) would seem to be 9 – 14 year$^{-1}$, corresponding to periods in the range 26 – 41 days.

We show spectrograms (again formed from 200 consecutive measurements) for $^{36}$Cl and $^{32}$Si data, for the frequency band 8 – 16 year$^{-1}$, in Figures 12 and 13, respectively. We see that both spectrograms give evidence of an oscillation with frequency of about 12.7 year$^{-1}$, which is compatible with a source in the solar radiative zone. For both spectrograms, this oscillation is evident over the time interval 1983.5 to 1985.5. However, the two spectrograms differ in that there is evidence of an oscillation with frequency about 11.2 year$^{-1}$ in the $^{36}$Cl spectrogram (Figure 12) but no evidence of a similar oscillation in the $^{32}$Si spectrogram (Figure 13). Hence, whatever the source of these oscillations, it does not have exactly the same influence on the $^{36}$Cl measurements and on the $^{32}$Si measurements. The depth of modulation of the oscillation at 12.7 year$^{-1}$ is 8.4 10$^{-4}$ for $^{36}$Cl and 7.6 10$^{-4}$ for $^{32}$Si.

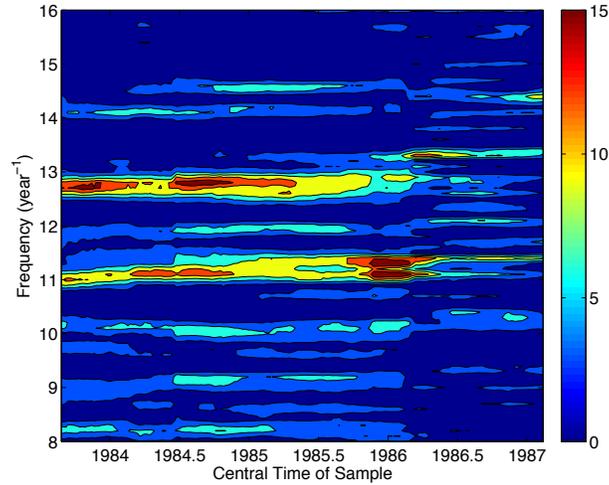

Figure 12. Spectrogram formed from $^{36}$Cl data for the frequency band 8 – 16 year$^{-1}$.

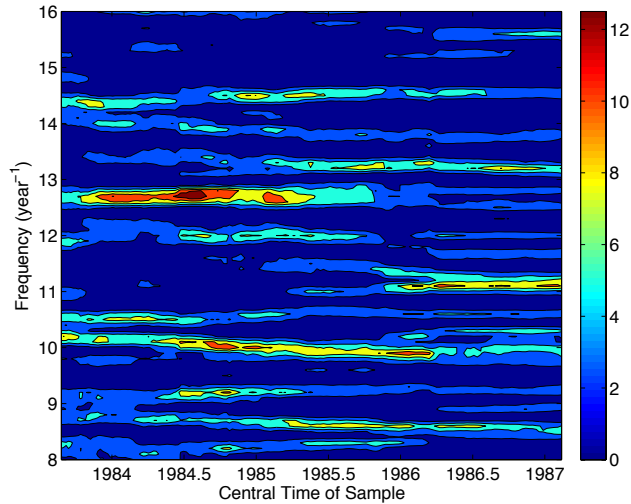

Figure 13. Spectrogram formed from $^{32}$Si data for the frequency band 8 – 16 year$^{-1}$.

In order to assess the significance of the features shown in Figures 12 and 13, it is necessary to estimate the probability of finding such a feature in a specified band. We proceed by first estimating the probability $P_1$ of finding a feature with the actual power in a band of unit width (1 year$^{-1}$). Since we plan to shuffle the data, the choice of band is unimportant – only the width matters. We see from Table 1 that the actual peak power in the band 12 – 13 year$^{-1}$ is 13.29 for the $^{36}$Cl power spectrum and we see from Table 2 that the actual peak power in the band 12 – 13 year$^{-1}$ is 8.69 for the $^{32}$Si power spectrum. The results of the shuffle tests are shown in Figures 14 and 15. We see from Figure 14 that there is a probability of only 7.4 10$^{-5}$ of finding by chance a feature of power 13.29 in a band of unit width by shuffling the $^{36}$Cl data. We see from Figure 15 that there is a probability of only 7.4 10$^{-3}$ of finding by chance a feature of power 8.69 in a band of unit width by shuffling the $^{32}$Si data.

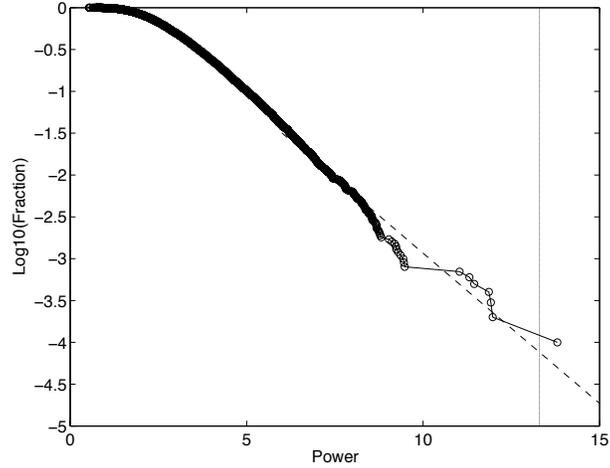

Figure 14. Results of a shuffle test of the rotational modulation in $^{36}$Cl data.
There is a probability of only 7.4 $10^{-5}$ of obtaining by chance
the actual power 13.29 in a band of unit width (1 year$^{-1}$).

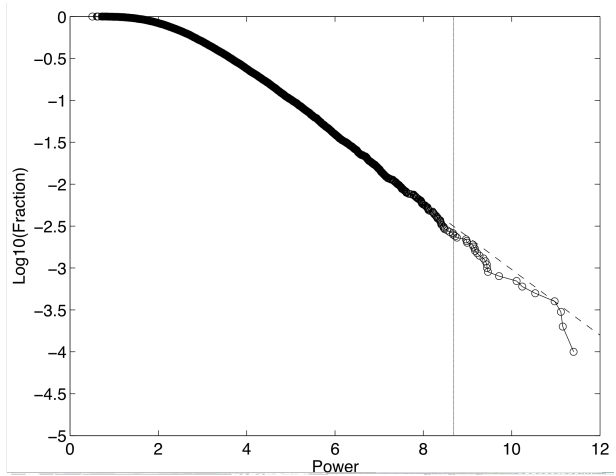

Figure 15. Results of a shuffle test of the rotational modulation in $^{32}$Si data.
There is a probability of only 3.2 $10^{-3}$ of obtaining by chance
the actual power 8.69 in a band of unit width (1 year$^{-1}$).

We may infer from these results the probabilities of finding such features in bands other than those of unit width. The probability of finding such a feature in a band of width $B$ is given by

$$P_B = 1 - (1 - P_1)^B . \qquad (4)$$

Hence, for the $^{36}$Cl data, if we choose to adopt a search band of 9 – 14 year$^{-1}$ (of width 5 year$^{-1}$) we arrive at a probability of 3.7 $10^{-4}$ of finding in that band a feature of power as large as or larger than the actual power (13.29). For the $^{32}$Si data, we would arrive at a probability of 0.016 of finding in that band a feature of power as large as or larger than the actual power (8.69). We find that the depths of modulation of the rotational signals shown in Figures 12 and 13 are 7 $10^{-4}$ for $^{36}$Cl and 6 $10^{-4}$ for $^{32}$Si.

**Section 5. Super-Kamiokande power spectrum analysis**
We now examine the Super-Kamiokande dataset as published in Yoo *et al.* (2003). This comprises 358 measurements of the B8 neutrino flux, of energy ≥ 5 MeV, acquired over the time interval June 2 1996 to July 15 2001. The mean flux over that time interval was 2.35 $10^6$ $cm^{-2}$ $s^{-1}$. The measurements, corrected for the varying Earth-Sun distance and normalized with respect to the mean flux, are shown in Figure 16.

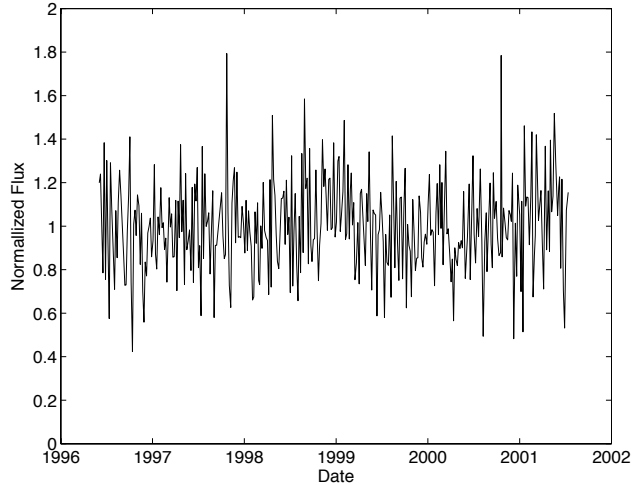

Figure 16. Neutrino flux measurements made by the Super-Kamiokande experiment (Yoo *et al.,* 2003), corrected for varying Earth-Sun distance, and normalized to mean value unity.

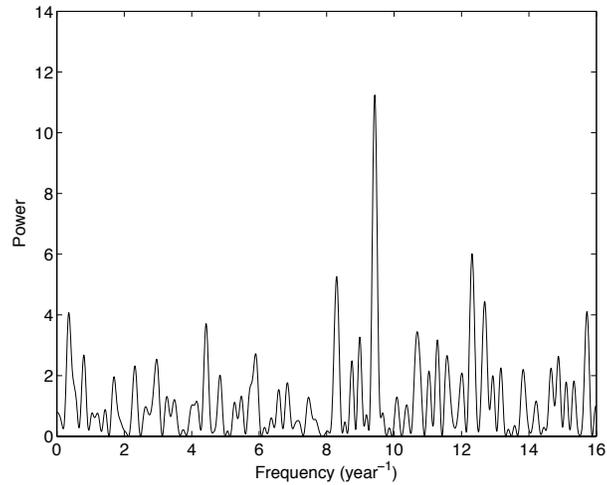

Figure 17. Power spectrum formed from Super-Kamiokande measurements, using a likelihood procedure that takes account of the start and end time of each run and of the error estimates.

Table 4. Top 20 peaks in the frequency band 0 – 16 $year^{-1}$ in the power spectrum formed from Super-Kamiokande $^8$B flux measurements.

| Frequency (year$^{-1}$) | Power | Order |
|---|---|---|
| 0.36 | 4.07 | 6 |
| 0.80 | 2.68 | 12 |
| 2.31 | 2.32 | 17 |
| 2.96 | 2.54 | 15 |
| 4.43 | 3.71 | 7 |
| 5.90 | 2.72 | 11 |
| 8.30 | 5.26 | 3 |
| 8.75 | 2.49 | 16 |
| 8.98 | 3.27 | 9 |
| 9.43 | 11.24 | 1 |
| 10.69 | 3.45 | 8 |
| 11.29 | 3.18 | 10 |
| 11.57 | 2.66 | 13 |
| 12.31 | 6.01 | 2 |
| 12.69 | 4.44 | 4 |
| 13.17 | 2.25 | 18 |
| 13.83 | 2.20 | 20 |
| 14.66 | 2.24 | 19 |
| 14.87 | 2.64 | 14 |
| 15.72 | 4.11 | 5 |

We have analyzed this dataset, using a likelihood procedure that takes account of the start time and end time and the symmetrized error estimate of each time bin (Sturrock, Caldwell, *et al.,* 2005). The resulting power spectrum is shown in Figure 17 and the top 20 peaks are shown in Table 4. We find the strongest peak at 9.43 year$^{-1}$ with power 11.24, and the second-strongest peak at 12.31 year$^{-1}$ with power 6.01.

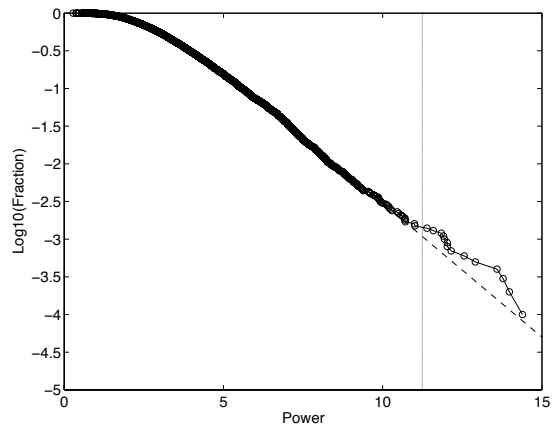

Figure 18. Results of a shuffle test of the rotational modulation in Super-Kamiokande data. There is a probability of only 10$^{-3}$ of obtaining by chance the actual power 11.24 or more in a band of unit width (1 year$^{-1}$).

We assess the significance of these oscillations by means of the shuffle test. Using the same procedure as in Section 4, we determine the maximum power in a band of unit width (1 year$^{-1}$) in each of 10,000 shuffles of the data. However, in analyzing Super-Kamiokande data we must organize the data into two sets: a timing set comprising the start and end time of each run, and a measurement set that comprises the flux measurement and the error estimate. The shuffle procedure does not change either set, but it re-assigns members of the timing set to members of the measurement set. The results of this shuffle procedure are shown in Figure 18. We find that there is a probability of only 0.001 of finding a peak of power 11.24 or more in unit bandwidth. This would lead to a probability of only 0.005 of finding a peak of power 11.24 or more in a rotational band extending from 9 to 14 year$^{-1}$.

This estimate agrees substantially with an analysis carried out by Ranucci (2006) who (assigning a mean time to each measurement bin) estimated a probability of 0.018 of finding such a peak in the frequency range 0 to 50 year$^{-1}$, which converts to a probability of 0.0004 for unit bandwidth.

We find from the shuffle test that there is a probability of 0.07 of finding a peak of power 6.01 or more in unit bandwidth. This would lead to a probability of 0.30 of finding such a peak in a rotational band extending from 9 to 14 year$^{-1}$. However, in view of the results of Section 4 shown in Figures 12 and 13, it is reasonable to consider also the probability of finding a peak of power 6.01 or more in the search band 12 – 13 year$^{-1}$, which is 0.07.

For comparison with the spectrograms formed from BNL data (Figures 8, 9, 12, 13), we now show the corresponding spectrograms formed from Super-Kamiokande data. Figure 19 shows the spectrogram formed (from measurements corrected for the varying Earth-Sun distance) for the frequency range 0 – 8 year$^{-1}$. We see that there is no evidence of an annual oscillation. Figure 20 shows the spectrogram formed for the frequency range 8 – 16 year$^{-1}$. We see evidence of a fairly steady feature at frequency 9.4 year$^{-1}$ and an intermittent feature centered approximately at 12.5 year$^{-1}$. It appears that the reason the 9.4 peak in the power spectrum is more significant than the 12.3 year$^{-1}$ peak is that the former oscillation remains fairly steady over the duration of the measurement whereas the latter is intermittent.

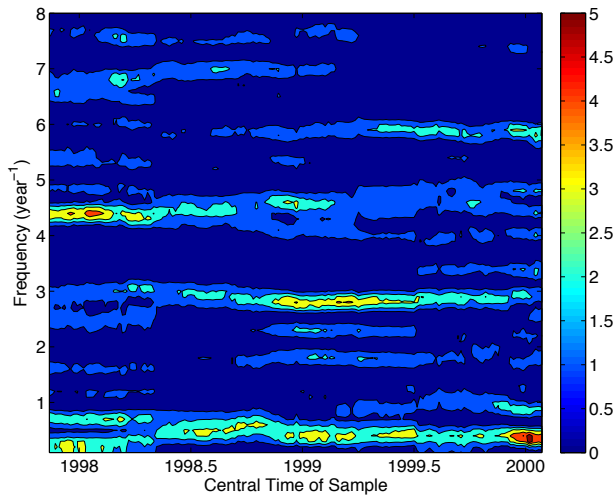

Figure 19. Spectrogram formed from Super-Kamiokande data (corrected for the varying Earth-Sun distance) for the frequency band 0 – 8 year$^{-1}$.

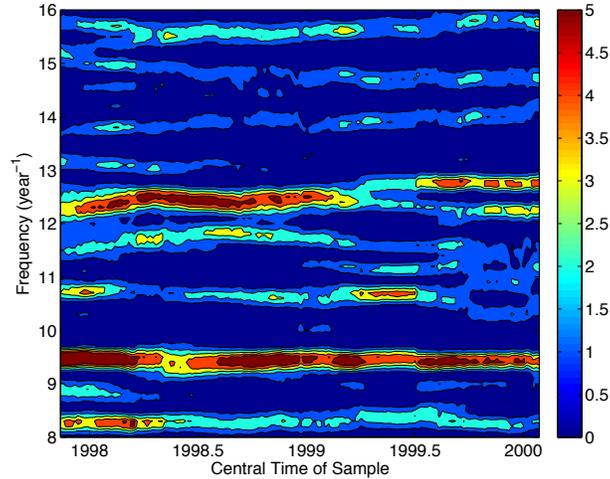

Figure 20. Spectrogram formed from Super-Kamiokande data
for the frequency band 8 – 16 year$^{-1}$.

For comparison with the depth of modulation of the rotational modulation found in BNL measurements, we note that the depth of modulation of the rotational modulation at 12.31 year$^{-1}$ in Super-Kamiokande data is 0.052.

We noted in Section 4 that the spectrogram formed from BNL $^{36}$Cl measurements shows a feature at 11 year$^{-1}$. There is weak evidence for a similar feature in Figure 20. These results raise the question – What pattern of internal rotation of the Sun might explain evidence for rotation features at 9.4 year$^{-1}$, 11 year$^{-1}$, and approximately 12.5 year$^{-1}$? We discuss this question in the next section.

**Section 6. Discussion**
Since an annual oscillation may be due to environmental influences on the experiment, the main focus of this article is whether or not beta-decay rates show evidence of an influence related to solar rotation, However, we shall first review the results of our search for an annual oscillation.

We found in Sections 2 and 3 that there is strong evidence for an annual oscillation in $^{36}$Cl measurements, but little evidence for such an oscillation in $^{32}$Si measurements. We estimated that the probability that the annual oscillation in $^{36}$Cl data is due to random fluctuations in the measurements is only 3 10$^{-7}$. We found no similar evidence of an annual oscillation in $^{32}$Si data. This is an interesting result. If the annual oscillation in the $^{36}$Cl data were due to an environmental influence such as temperature fluctuations, one would expect to see similar evidence in both datasets since both nuclides were subject to exactly the same influences. Hence the fact that we see strong evidence of an annual oscillation in $^{36}$Cl data but not in $^2$Si data suggests that the annual oscillation evident in $^{36}$Cl data is not due to an environmental influence such as temperature.

Moreover the spectrogram shown in Figure 8 shows that the annual oscillation in $^{36}$Cl data is intermittent rather than steady and we find (Table 3 and Figure 10) that the annual oscillation has its maximum at phase 0.7, i.e. in mid-September. These results argue against the early suggestion (Falkenberg 2001; Jenkins *et al.,* 2009) that an annual oscillation in a decay rate may be due to the annual oscillation of a solar flux due to the annual variation of the Earth-Sun distance, since this influence would be steady and would lead to an amplitude of 3.3% and a phase of maximum near January 1.

These results therefore appear to be incompatible with the hypothesis that the annual oscillations are due to the influence of a steady flux of solar neutrinos. However, they appear to be compatible with the hypothesis that the annual oscillations are due to the influence of cosmic neutrinos if one allows for the possibility that different nuclides are responsive to neutrinos of different energy and that the spatial structure and energy composition of cosmic neutrinos are variable. This hypothesized variability in the structure and composition of the cosmic-neutrino distribution may be the result of stochastic deflections and stochastic acceleration due to gravitational encounters with stars, galactic nuclei, black holes, etc., processes that would depend on the speed (and therefore the energy) of the particles.

We now consider the evidence for an influence of solar rotational modulation on the beta-decay rates analyzed in Section 4. We found evidence for a transient oscillation of frequency about 12.7 year$^{-1}$ in both $^{36}$Cl and $^{32}$Si spectrograms (Figures 12 and 13). This frequency is compatible with an influence on the solar neutrino flux that occurs in the radiative zone. We found from the shuffle test that these features could have occurred by chance in a band of unit width with probabilities 7.4 10$^{-5}$ and 3.2 10$^{-3}$, respectively. For a search band of width 5 year$^{-1}$, the probability of finding both features within a sub-band of 1 year$^{-1}$ is 1.2 10$^{-6}$.

Figure 12 shows a feature with frequency about 11.0 year$^{-1}$ early in the spectrogram formed from $^{36}$Cl data. There is a hint of a similar feature late in the spectrogram formed from $^{32}$Si data (Figure 13). A synodic frequency of 11 year$^{-1}$ corresponds to a sidereal frequency of 12 year$^{-1}$, which is the rotation frequency of a region (presumed to be a tachocline) that appears to be the site of r-mode oscillations in the Sun (Sturrock, 2008; Sturrock & Bertello, 2010; Sturrock, Fischbach, *et al.,* 2011; Sturrock, Parkhomov, *et al.,* 2012; Sturrock, Bertello, *et al.,* 2013; Sturrock, Bush, *et al.,* 2015).

Figure 13 also shows evidence of a transient feature with a frequency of about 10 year$^{-1}$ in $^{32}$Si data. We have speculated that the oscillation with a frequency of about 9.4 year$^{-1}$ found in Super-Kamiokande data (evident in Figures 17 and 20) may have its origin in the solar core (Sturrock, Parkhomov, *et al.,* 2012). Hence an oscillation with a frequency of about 10 year$^{-1}$ may have its origin in a region between the core and the inner tachocline.

Figure 21 gives a schematic representation of the sidereal internal rotation of the Sun, as inferred from studies of rotational and r-mode oscillations. (Note that the synodic rotation rate is smaller than the sidereal rate by 1 year$^{-1}$.) As suggested in Section 4, it seems most likely that rotational modulation may be attributed to the RSFP process (Akhmedov, 1988). Modulation at about 9.5 year$^{-1}$, evident in Super-Kamiokande data (Figure 20), probably occurs in or near the core (Sturrock, Parkhomov, *et al.,* 2012). Modulation at about 11 year$^{-1}$, evident in $^{36}$Cl data (Figure 12) and weakly evident in both $^{32}$Si data (Figure 13) and Super-Kamioknde data (Figure 20), may be attributed to the inner tachocline. Modulation in the range 12 – 13 year$^{-1}$, evident in both $^{36}$Cl and $^{32}$Si data (Figures 12 and 13) and in Super-Kamiokande data (Figure 20), probably occurs in the radiative zone. Modulation at about 10 year$^{-1}$, weakly evident in $^{32}$Si data (Figures 13), may occur in an outer region of the core, or an inner region of the inner tachocline.

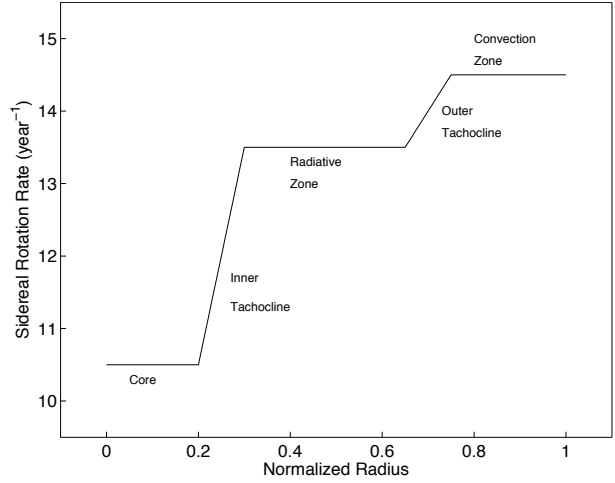

Figure 21. Equatorial cut of the conjectured internal rotation of the Sun.

It is interesting to compare the oscillations with frequencies in the range 12 – 13 year$^{-1}$ evident in both $^{36}$Cl and $^{32}$Si data (Figures 12 and 13) and the oscillation in the same range evident in Super-Kamiokande data (Figure 20). The two datasets were not acquired in the same time interval. However, we may consider the possibility that both datasets are typical of what might be obtained from beta-decay and neutrino measurements and explore the implications of that assumption.

The Super-Kamiokande experiment is sensitive to neutrinos of energy above 5 MeV. We see from Bahcall (1989, p. 154) that these are the $^8$B neutrinos and that, over the band 5 – 9 MeV, the flux at Earth is given by

$$F(^{8}B) = 10^{6.6} \text{ cm}^{-2}\text{ s}^{-1} . \qquad (5)$$

We found in Section 5 that the depth of modulation of the signal detected by Super-Kamiokande is given by

$$\Delta(^{8}B) = 10^{-1.3} . \qquad (6)$$

Hence the rotational signal in the Super-Kamiokande data can be attributed to an equivalent flux given by

$$\Delta F(^{8}B) = \Delta(^{8}B) \times F(^{8}B) = 10^{5.3} \text{ cm}^{-2}\text{ s}^{-1} . \qquad (7)$$

The half lives of $^{36}$Cl and $^{32}$Si are

$$T_{1/2}(^{36}Cl) = 300,000 \text{ y and } T_{1/2}(^{32}Si) = 172 \text{ y} , \qquad (8)$$

which convert to decay rates

$$\Gamma(^{36}Cl) = 10^{-13.1} \text{ s}^{-1} \text{ and } \Gamma(^{32}Si) = 10^{-9.9} \text{ s}^{-1} . \qquad (9)$$

We found in Section 5 that the depth of modulation of the signals detected by BNL are given by

$$\Delta(^{36}\text{Cl}) = 7\,10^{-4} \text{ and } \Delta(^{32}\text{Si}) = 6\,10^{-4} \ . \tag{10}$$

Hence the amplitudes of the rotational modulations of the decay rates are given by

$$\Lambda\Gamma(^{36}\text{Cl}) = \Lambda(^{36}\text{Cl}) \times \Gamma(^{36}\text{Cl}) = 10^{-16.3}\ \text{s}^{-1}, \text{ and}$$
$$\Lambda\Gamma(^{32}\text{Si}) = \Lambda(^{32}\text{Si}) \times \Gamma(^{32}\text{Si}) = 10^{-13.1}\ \text{s}^{-1}\ . \tag{11}$$

If we now define an equivalent cross section $\sigma$ as follows

$$\Delta\Gamma = \sigma \times \Delta F\ , \tag{12}$$

we obtain the estimates

$$\sigma(^{36}\text{Cl},^{8}\text{B}) = 10^{-21.6}\ \text{cm}^{-2} \text{ and } \sigma(^{32}\text{Si},^{8}\text{B}) = 10^{-18.4}\ \text{cm}^{-2}\ . \tag{13}$$

We include the names of both the nuclide and the neutrino type to emphasize that the cross sections have been estimated only for those specific combinations.

Using these estimates of the cross sections, it is interesting to calculate the neutrino flux that would be required to explain the decay rates. This is found to be $10^{8.8}$ cm$^{-2}$ s$^{-1}$ for each nuclide. It is also interesting to estimate the equivalent cross-section of Super-Kamiokande. Hosaka *et al.* (2006) give (for Kamiokande I) the total count as 22,404 and the total live time as 1496 days, which lead to a count rate of $10^{-3.76}$ s$^{-1}$. The fiducial mass is given as 22.5 kt, which will contain $10^{33.17}$ electrons. Hence the event rate per electron is $10^{-36.93}$ s$^{-1}$. Since the $^{8}$B neutrino flux is $10^{6.6}$ cm$^{-2}$ s$^{-1}$, we may estimate the cross section per electron to be $10^{-43.5}$ cm$^2$, which is close to the theoretical value $10^{-43.2}$ cm$^2$ (Bahcall, 1989). Hence the total cross section of Super-Kamiokande is $10^{-10.4}$ cm$^2$, which is the cross section of approximately $10^8$ atoms of $^{32}$Si. Hence 1 pg of $^{32}$Si would have a capture rate of $^{8}$B neutrinos that is 200 times the capture rate of Super-Kamiokande.

As a typical "exempt" source of low activity, we may consider a $10\,\mu$Ci specimen of $^{32}$Si, which would generate $10^{5.6}$ counts per second. Hence a one-day measurement would accumulate N = $10^{10.6}$ counts. The statistical uncertainty of this measurement would be N$^{-1/2}$, i.e. $10^{-5.3}$. By contrast, Super-Kamiokande registers 15 counts per day, for which the statistical uncertainty is 0.26 ($10^{-0.6}$), a difference of almost $10^5$.

There is at this time no theory that can explain the apparent influence of neutrinos on beta decays. The Fermi theory of beta decays (Griffiths, 1987), which is well established experimentally, must remain a key part of a theory. However, something else is required to explain the fact that decay rates are variable, and that the variability is associated with neutrinos. According to our analysis of GSI (Geological Survey of Israel) data, decay products tend to move in the direction of the incoming neutrinos (Sturrock, Fischbach, *et al.*, 2015). These results suggest that although the Fermi model may primarily determine the probability of the occurrence of a beta decay, some aspects of the decay may be influenced by the ambient neutrinos. This is reminiscent of processes in a plasma: the force between two charged particle is due to the electrical charges on the particles, but the force is modified by the collective behavior (known as *Debye shielding*) of the surrounding electron distribution (Sturrock, 1994). Fluctuations in the plasma will lead to fluctuations in the force that one particle exerts on another particle. This collective behavior is due to the long-range electrostatic force. Perhaps there is a currently unknown long-range force that couples neutrinos to neutrinos.

A procedure that involves two separate mechanisms, such as the one proposed, could explain why nuclides with very different decay rates nevertheless exhibit oscillations with much the same degree of modulation.

This influence may effectively be manifested in only a small energy range of the decay process. It may therefore be relevant to note that early reports from the Troitsk collaboration (Lobashev *et al.,* 1999, 2001) indicate that measurements of the decay rate of tritium are suggestive of variability with an annual and/or a semi-annual period in the high-energy range of the decay products. Aseev *et al. (*2012) do not find this variability in a smaller sample of Troitsk data, but Falkenberg (2001) has claimed to detect an annual oscillation, and Vevprev *et al.* (2012) have claimed to detect diurnal and 27-day oscillations, in tritium decay measurements.

If beta decays are indeed influenced by neutrinos, beta-decay experiments may prove to be useful detectors not only of solar neutrinos but also of cosmic neutrinos and possibly of supernova neutrinos.

We thank D.E. Alburger, G. Harbottle and E.F. Norton for providing us with data and other information concerning their experiment at the Brookhaven National Laboratory, and we thank the Super-Kamiokande Consortium for publishing the results of the Super-Kamiokande I Experiment.